\documentclass[aps,twocolumn,nofootinbib,groupedaddress]{revtex4-1}

\usepackage[english]{babel}
\usepackage{graphicx}
\usepackage{amsmath,amssymb,amsbsy,amstext, amsthm, simplewick, amsfonts}
\usepackage{dcolumn}

\usepackage{hyperref}
\hypersetup{
    colorlinks=true,
    linkcolor=purple,
    filecolor=purple,      
    urlcolor=purple,
    citecolor=purple
    }
    
\usepackage{comment}
\usepackage{bm}
\usepackage{xcolor}
\usepackage{braket}
\usepackage{tensor}
\usepackage{ulem}

\newcommand{\dd}{\mathrm{d}}

\def\Mp{M_{\rm pl}}

\usepackage{colortbl}
\definecolor{pyblue}{RGB}{31, 119, 180}
\definecolor{pyred}{RGB}{214, 39, 40}
\definecolor{pygreen}{RGB}{44, 160, 44}

\begin{document}


\title{Cosmological Flow of Primordial Correlators}

\author{Denis Werth,$^1$ Lucas Pinol,$^2$ and Sébastien Renaux-Petel$^1$}

\affiliation{$^1$Sorbonne Université, Institut d'Astrophysique de Paris, CNRS, UMR 7095, 98 bis bd Arago, 75014 Paris, France \\
$^2$Instituto de Física Teórica UAM-CSIC, c/ Nicolás Cabrera 13-15, 28049 Madrid, Spain}

\begin{abstract}
Correlation functions of primordial density fluctuations provide an exciting probe of the physics governing the earliest moments of our Universe. However, the standard approach to compute them is technically challenging. Theoretical predictions are therefore available only in restricted classes of theories. In this Letter, we present a complete method to systematically compute tree-level inflationary correlators. This method is based on following the time evolution of equal-time correlators and it accurately captures all physical effects in any theory. These theories are conveniently formulated at the level of inflationary fluctuations, and can feature any number of degrees of freedom with arbitrary dispersion relations and masses, coupled through any type of time-dependent interactions. We demonstrate the power of this approach by exploring the properties of the cosmological collider signal, a discovery channel for new high-energy physics, in theories with strong mixing and in the presence of features. This work lays the foundation for a universal program to  assist our theoretical understanding of inflationary physics and generate theoretical data for an unbiased interpretation of upcoming cosmological observations. 
\end{abstract}

\maketitle


\paragraph*{\bf Introduction.} 
Cosmology is about understanding time. Indeed, the physics governing the Universe is deciphered through the \textit{time evolution} of density perturbations, from the beginning of the hot big bang to the late-time galaxy distribution. Remarkably, it is believed that these perturbations have emerged from  quantum zero-point fluctuations during a period of accelerated expansion \cite{Guth:1980zm, Linde:1981mu, Albrecht:1982wi}, i.e., cosmic inflation, providing the initial seeds for the subsequent evolution of cosmological structures \cite{Mukhanov:1981xt, Starobinsky:1982ee, Hawking:1982cz, Guth:1982ec, Bardeen:1983qw}. Tracking down the cosmological flow of inflationary fluctuations thus connects the quantum laws of physics at a fundamental level to the largest observable scales.

Yet, the correct theory of inflation remains unknown, and a central challenge is to decode it through the study of inflationary correlators, namely spatial correlation functions of the curvature perturbation field $\zeta(\bm{x}, t)$. On large scales, current data have already  well constrained the physics of inflation, especially at the linear level~\cite{Planck:2018jri}. However, much information---e.g.~the number of fields active during inflation, together with their mass spectra, spins, sound speeds, and how they interact---is encoded in nonlinearities 
(see, e.g., \cite{Achucarro:2022qrl} for a recent review), of which the primary observable is the three-point correlator (bispectrum) (see~\cite{Planck:2019kim} for current bounds).

Computing equal-time correlators given an inflationary theory is a well-established procedure~\cite{Weinberg:2005vy}. From first principles, calculations can be carried to arbitrary orders of perturbation theory. However, this program hides a daunting complexity: perturbation theory becomes intractable for realistic situations. 
The root of this difficulty resides in the challenge to track the detailed time evolution of the physics in the bulk of spacetime. Consequently, for technical reasons, most of the theoretical predictions have been derived under stringent assumptions, such as assuming weak mixing, perfect (or almost) scale invariance, large hierarchy of masses and couplings, and considering single-exchange diagrams~\cite{Achucarro:2022qrl}. Therefore, they do not cover the vast panorama of inflationary scenarios \cite{Baumann:2014nda}. This can completely bias our interpretation of data and reveals the need to develop an approach that makes accurate predictions for \textit{all} physically motivated inflationary theories. 

 \begin{figure}[b!]
       \centering    \includegraphics[width=0.75\columnwidth,trim={0pt 6pt 0pt 0}]{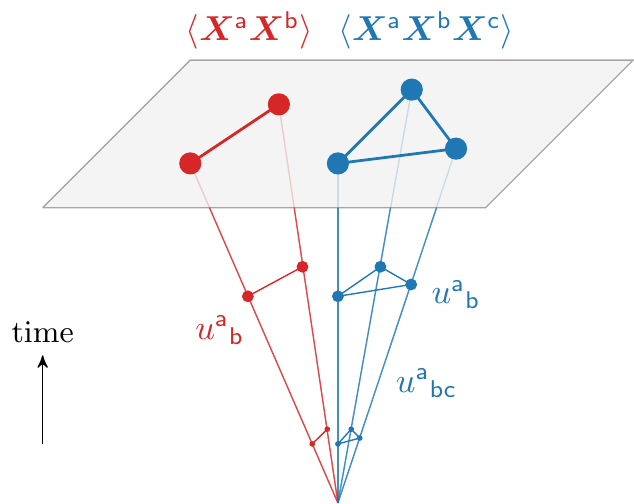}
       \caption{Schematic diagram of the cosmological flow. The time evolution of inflationary correlators, given by Eqs.~(\ref{eq: 2pt time evolution},~\ref{eq: 3pt time evolution}), is tracked from initial quantum fluctuations in the infinite past to the end of inflation. All the theory dependence is left in the precise form of the tensors $\tensor{u}{^{\sf{a}}}{_{\sf{b}}}$ and $\tensor{u}{^{\sf{a}}}{_{\sf{bc}}}$ driving this flow.}
        \label{fig: The Cosmological Flow} 
 \end{figure}

In this Letter and its companion paper~\cite{CosmoFlow}, we present the \textit{cosmological flow}, a systematic framework to compute primordial correlators. It is based on solving differential equations in time to track the evolution of primordial correlators through the entire spacetime during inflation (see Fig.~\ref{fig: The Cosmological Flow}), for theories formulated straight at the level of fluctuations. In contrast to previous works~\cite{Seery:2012vj, Mulryne:2013uka, Dias:2016rjq, Mulryne:2016mzv, Butchers:2018hds}, our approach is not limited to nonlinear sigma models of multi-field inflation.
Instead, by adopting a philosophy that focuses on the study of correlators through effective field theories (EFT) of inflationary fluctuations~\cite{Creminelli:2006xe,Cheung:2007st, Senatore:2010wk,Noumi:2012vr,Pinol:2024arz}, we are able to describe any theory. 
Our formalism yields exact tree-level results in theories featuring an arbitrary number of fluctuating degrees of freedom with varied (possibly nonlinear) dispersion relations and masses, coupled through any type of time-dependent interactions. It also solves for correlators directly connected to observables without relying on specific background-field dynamics, making the physical effects of symmetries explicit. 

This work offers new possibilities for the study of inflationary correlators. One of the motivations for studying them lies in the appreciation that inflation is a one-of-a-kind window to probe fundamental physics at the highest reachable energies, comparable to the Hubble scale $H$ during inflation, which can be as high as $10^{14}$~GeV. Information about new physics---e.g., the presence of heavy particles of masses $m\sim H$---can be inferred from oscillatory patterns (or specific power law behaviors) present in the squeezed limit of the bispectrum \cite{Chen:2009zp, Noumi:2012vr, Arkani-Hamed:2015bza}.\footnote{More generally, smoking guns of new physics are encoded in particular soft limits of higher-order inflationary correlators.} Our approach provides a systematic way to explore the characteristics of this cosmological collider signal in theories that are difficult to grasp analytically.
First, as a heavy field weakly mixed with the curvature perturbation leads to the same cosmological collider frequency as a light but strongly mixed one, we show how complete predictions can be straightforwardly obtained in order to break theoretical degeneracies. 
Second, we demonstrate that, due to the universal nonlinearly realized boost symmetry, a time-dependent mixing leads to cosmological collider signals composed of modulated frequencies.

These results contradict some commonly held beliefs that the detection of such signals pinpoints the mass of a new particle. This highlights the importance of thoroughly studying signatures of all early-universe theories, as the cosmological flow approach enables one to do, for correctly interpreting cosmological data. More generally, our work paves the way for a far-reaching program of studying the phenomenology of inflationary correlators, shifting our focus from technical considerations to the unbiased exploration of the rich physics of inflation.


\vskip 4pt
\paragraph*{\bf Primordial Fluctuations.} To begin, let us consider a set of scalar fluctuations $\bm{\varphi}^\alpha (\bm{x}, t)$, and $\bm{p}^\beta(\bm{x}, t)$ the corresponding conjugate momenta. For practical purposes, we gather these fields and momenta in a phase-space vector $\bm{X}^a \equiv (\bm{\varphi}^\alpha, \bm{p}^\beta)$.\footnote{Following the conventions of~\cite{Dias:2016rjq}, Greek indices $\alpha, \beta, \gamma, \ldots$ run over fields, and Latin indices $a, b, c, \dots$ run over phase-space coordinates, organized so that a block of field labels is followed by a block of momentum labels, in the same order.} We denote the corresponding operators in Fourier space with \textsf{sans serif} indices $\bm{X}^{\sf{a}}(\bm{k}, t)$. They verify the canonical commutation algebra $[\bm{X}^{\sf{a}}, \bm{X}^{\sf{b}}] = i \epsilon^{\sf{a}\sf{b}} \equiv i \,(2\pi)^3\delta^{(3)}(\bm{k}_a + \bm{k}_b)\,\epsilon^{ab}$ with
the $2N\times2N$ matrix written in block form
$\epsilon^{ab} \equiv 
\begin{pmatrix}
\bf{0} & \bf{1} \\
\bf{-1} & \bf{0}
\end{pmatrix}$.
These inflationary fluctuations are described by a Hamiltonian $H(\bm{\varphi}^{\alpha}, \bm{p}^\beta)$ which is a functional of the phase-space coordinates. 
It takes the form of a series expansion in powers of fluctuations
\begin{equation}
\label{eq: Hamiltonian}
    H = \frac{1}{2!}H_{\sf{ab}} \bm{X}^{\sf{a}} \bm{X}^{\sf{b}} + \frac{1}{3!}H_{\sf{abc}} \bm{X}^{\sf{a}} \bm{X}^{\sf{b}} \bm{X}^{\sf{c}} + \ldots\,,
\end{equation}
where we adopt the extended Fourier summation convention for repeated indices to indicate a sum including integrals over Fourier modes. The tensors $H_{\sf{ab}}, H_{\sf{abc}}, \ldots$, which can be taken symmetric without loss of generality, are arbitrary functions of time and momenta. This form of the Hamiltonian is completely general and captures \textit{all} theories of scalar fluctuations.
The fully nonlinear equations of motion read
\begin{equation}
\label{eq: Heisenberg EOM}
\begin{aligned}
    \frac{\dd \bm{X}^{\sf{a}}}{\dd t} &= i\,[H, \bm{X}^{\sf{a}}] \\
    &= \epsilon^{\sf{ac}}H_{\sf{cb}}\bm{X}^{\sf{b}} + \frac{1}{2!}\epsilon^{\sf{ad}}H_{\sf{dbc}}\bm{X}^{\sf{b}}\bm{X}^{\sf{c}} + \dots \\
    &= \tensor{u}{^{\sf{a}}}{_{\sf{b}}}\bm{X}^{\sf{b}} + \frac{1}{2!} \tensor{u}{^{\sf{a}}}{_{\sf{bc}}}\bm{X}^{\sf{b}}\bm{X}^{\sf{c}} + \ldots \,,
\end{aligned}
\end{equation}
where the third line defines the tensors $\tensor{u}{^{\sf{a}}}{_{\sf{b}}}, \tensor{u}{^{\sf{a}}}{_{\sf{bc}}}, \dots$. Written in this form, it is clear that Eq.~(\ref{eq: Heisenberg EOM}) encodes both the full classical evolution of $\bm{X}^{\sf{a}}$ \textit{and} their quantum properties. 


\vskip 4pt
\paragraph*{\bf Time Evolution of Primordial Correlators.} We are interested in \textit{equal-time} correlators of composite operators $\bra{\Omega} \mathcal{O}(\bm{X}^{\sf{a}}) \ket{\Omega}$ evaluated at time $t$, where $\ket{\Omega}$ is the vacuum of the full interacting theory. Because the dynamics governed by Eq.~(\ref{eq: Heisenberg EOM}) cannot be solved exactly in full generality, we choose the quadratic Hamiltonian $H_0 = \frac{1}{2!}H_{\sf{ab}}\bm{X}^{\sf{a}}\bm{X}^{\sf{b}}$ to evolve the interaction-picture operators defined by $X^{\sf{a}} \equiv \mathcal{U}^\dagger \bm{X}^{\sf{a}}\, \mathcal{U}$, thus resorting to a perturbative description of the interactions encoded in $H_{\text{I}} = H - H_0$ \cite{Weinberg:1995mt, Peskin:1995ev}. In this way, equal-time correlators are given by the well-known in-in formula \cite{Weinberg:2005vy}
\begin{equation}
\label{eq: Interaction-picture correlators}
    \bra{\Omega} \mathcal{O}(\bm{X}^{\sf{a}})\ket{\Omega} = \bra{0} \mathcal{U}\,\mathcal{O}(X^{\sf{a}})\,\mathcal{U}^\dagger\ket{0}\,,
\end{equation}
where $\mathcal{U} = \bar{\text{T}} \exp [i\int_{-\infty(1-i\epsilon)}^t H_{\text{I}}(t')\,\dd t']$
with $\bar{\text{T}} = \text{T}^\dagger$ the anti-time ordering operator, and $\ket{0}$ is the vacuum of the free theory. The $X^{\sf{a}}$ evolve with the \textit{full} quadratic Hamiltonian. Working at tree level and up to three-point correlators, one can expand the exponentials in Eq.~(\ref{eq: Interaction-picture correlators}) to obtain
\begin{gather}
\label{eq: Tree-level correlators}
    \begin{aligned}
    \langle\bm{X}^{\sf{a}} \bm{X}^{\sf{b}} \rangle &= \bra{0}X^{\sf{a}}X^{\sf{b}}\ket{0}\,,\\
    \langle\bm{X}^{\sf{a}} \bm{X}^{\sf{b}}\bm{X}^{\sf{c}} \rangle &= \bra{0} \frac{i}{3!}\int_{-\infty}^t\dd t' H_{\mathsf{def}}\left[X^{\mathsf{d}}X^{\mathsf{e}}X^{\mathsf{f}}, X^{\mathsf{a}}X^{\mathsf{b}}X^{\mathsf{c}} \right]  \ket{0}\,.
    \end{aligned}
\raisetag{42pt}
\end{gather}
For the three-point correlators, external operators are evaluated at the time $t$, and internal operators at time $t'$. The simplicity of the cosmological flow lies in the ability to find, from first principles, a closed system of differential equations in time at the level of correlators. 
Indeed, differentiating Eq.~(\ref{eq: Tree-level correlators}) with respect to time $t$ and using the equations of motion for the interaction-picture fields $X^{\sf{a}}$, the cosmological flow is given by
\begin{subequations}
\begin{align}
    \frac{\dd}{\dd t} \langle \bm{X}^{\sf{a}} &\bm{X}^{\sf{b}} \rangle = \tensor{u}{^{\sf{a}}}{_{\sf{c}}} \langle \bm{X}^{\sf{c}} \bm{X}^{\sf{b}} \rangle + \tensor{u}{^{\sf{b}}}{_{\sf{c}}} \langle \bm{X}^{\sf{a}} \bm{X}^{\sf{c}} \rangle\,,\label{eq: 2pt time evolution}\\
    \begin{split}
    \frac{\dd}{\dd t} \langle \bm{X}^{\sf{a}} \bm{X}^{\sf{b}}&\bm{X}^{\sf{c}} \rangle = \tensor{u}{^{\sf{a}}}{_{\sf{d}}} \langle \bm{X}^{\sf{d}} \bm{X}^{\sf{b}}\bm{X}^{\sf{c}} \rangle \\
    &+ \tensor{u}{^{\sf{a}}}{_{\sf{de}}}\langle \bm{X}^{\sf{b}} \bm{X}^{\sf{d}} \rangle\langle \bm{X}^{\sf{c}} \bm{X}^{\sf{e}} \rangle + (2\text{ perms})\,.\label{eq: 3pt time evolution}
    \end{split}
\end{align}
\end{subequations}
Equation (\ref{eq: 2pt time evolution}) couples all two-point correlators, 
including those which contain conjugate momenta, and correctly captures \textit{all} physical effects arising from quadratic operators in the theory.
The structure of Eq.~(\ref{eq: 3pt time evolution}) allows the flow of each kinematical configuration to be tracked separately.
Considering the Bunch-Davies state in flat FLRW backgrounds,
as we do in the following, the initial conditions for Eqs.~(\ref{eq: 2pt time evolution}, \ref{eq: 3pt time evolution}) can be readily derived analytically provided one initializes the correlators sufficiently in the far past. Naturally, the method equally works for any other state and is not restricted to inflation.


\vskip 4pt
\paragraph*{\bf Goldstone Description.} 
Our EFT-based formalism, considering straight fluctuations around FLRW backgrounds, enables us to directly solve for correlators connected to observables.
The spontaneous breaking of boost symmetry in cosmological backgrounds implies the unavoidable presence of a (canonically normalized) Goldstone boson $\pi_c(\bm{x}, t)$ describing adiabatic fluctuations  \cite{Creminelli:2006xe, Cheung:2007st}. At linear order, the field $\pi_c$ is related to the curvature perturbation by $\zeta = -H c_s^{3/2} f_\pi^{-2}\pi_c$ where $c_s$ is the propagation speed of $\pi_c$ and $f_\pi^4 \equiv 2\Mp |\dot{H}|c_s$ is the symmetry breaking scale. 
Based on a systematic classification of unitary gauge operators and nonlinearly realized symmetries, let us now consider an additional relativistic massive scalar field $\sigma(\bm{x}, t)$ with mass $m$, coupled to $\pi_c$ through the following interacting Lagrangian:
\begin{equation}
\label{eq: pi-sigma theory}
\begin{aligned}
    \mathcal{L}/a^3 &= \rho \dot{\pi}_c\sigma + c_s^{3/2}\frac{\rho}{2f_\pi^2} \frac{(\partial_i \pi_c)^2}{a^2}\sigma + c_s^{3/2} \frac{\dot{\rho}}{f_\pi^2}\,\pi_c\dot{\pi}_c\sigma\\
    &- \frac{1}{2\Lambda}\dot{\pi}_c^2\sigma - \frac{1}{2}\alpha\dot{\pi}_c \sigma^2 -\mu \sigma^3\,,
\end{aligned}
\end{equation}
where $\rho, \Lambda, \alpha$ and $\mu$ are---in general time-dependent---couplings. For the purpose of focusing on mixing interactions, we have omitted ever-present self-interactions of $\pi_c$ and have taken the decoupling limit where gravitational interactions vanish. Note that there is no \textit{a priori} model-building requirement on the size of the dimensionless quadratic coupling $\rho/H$, and we allow this mixing parameter to be of the order of one or larger, yet still evading strong coupling, which we call the strong mixing regime. A universal aspect, which does not rely on particular (classes of) models, is that this coupling $\rho$ fixes both the quadratic interaction $\dot{\pi}_c \sigma$ and the cubic interactions $(\partial_i \pi_c)^2\sigma$ and $\pi_c\dot{\pi}_c\sigma$. This is a consequence of $\pi_c$ nonlinearly realizing time diffeomorphisms, as these interactions are generated by the same operator $\rho(t)\delta g^{00}\sigma$ in the unitary gauge, after reintroducing the Goldstone boson $t \rightarrow t+\pi$. After performing a Legendre transform, the found Hamiltonian can be arranged in the form (\ref{eq: Hamiltonian}) in terms of the phase-space vector $\bm{X}^a = (\pi_c, \sigma, p_{\pi}, p_\sigma)$. 
The identification of the quadratic tensor $\tensor{u}{^{\sf{a}}}{_{\sf{b}}} \equiv (2\pi)^3 \delta^{(3)}(\bm{k}_a - \bm{k}_b) \, \tensor{u}{^{a}}{_{b}}$, defined in Eq.~(\ref{eq: Heisenberg EOM}), follows simply, yielding
\begin{equation}
    \tensor{u}{^{a}}{_{b}} = 
    \begin{bmatrix}
        0 & -\rho & 1 & 0\\
        0 & 0 & 0 & 1 \\
        -c_s^2 k_a^2/a^2 & 0 & 0 & \rho \\
        0 & -\left(k_a^2/a^2+m^2+\rho^2\right) & 0 & 0
    \end{bmatrix}\,,
\end{equation}
and likewise for the cubic tensor $\tensor{u}{^{\sf{a}}}{_{\sf{bc}}}$ whose precise form can be found in~\cite{CosmoFlow}. We now apply our formalism to the study of cosmological collider signals, crucially focusing on the symmetries revealed by the EFT perspective.

\begin{figure}[t!]
    \centering    \includegraphics[width=1\columnwidth,trim={0pt 6pt 0pt 0}]{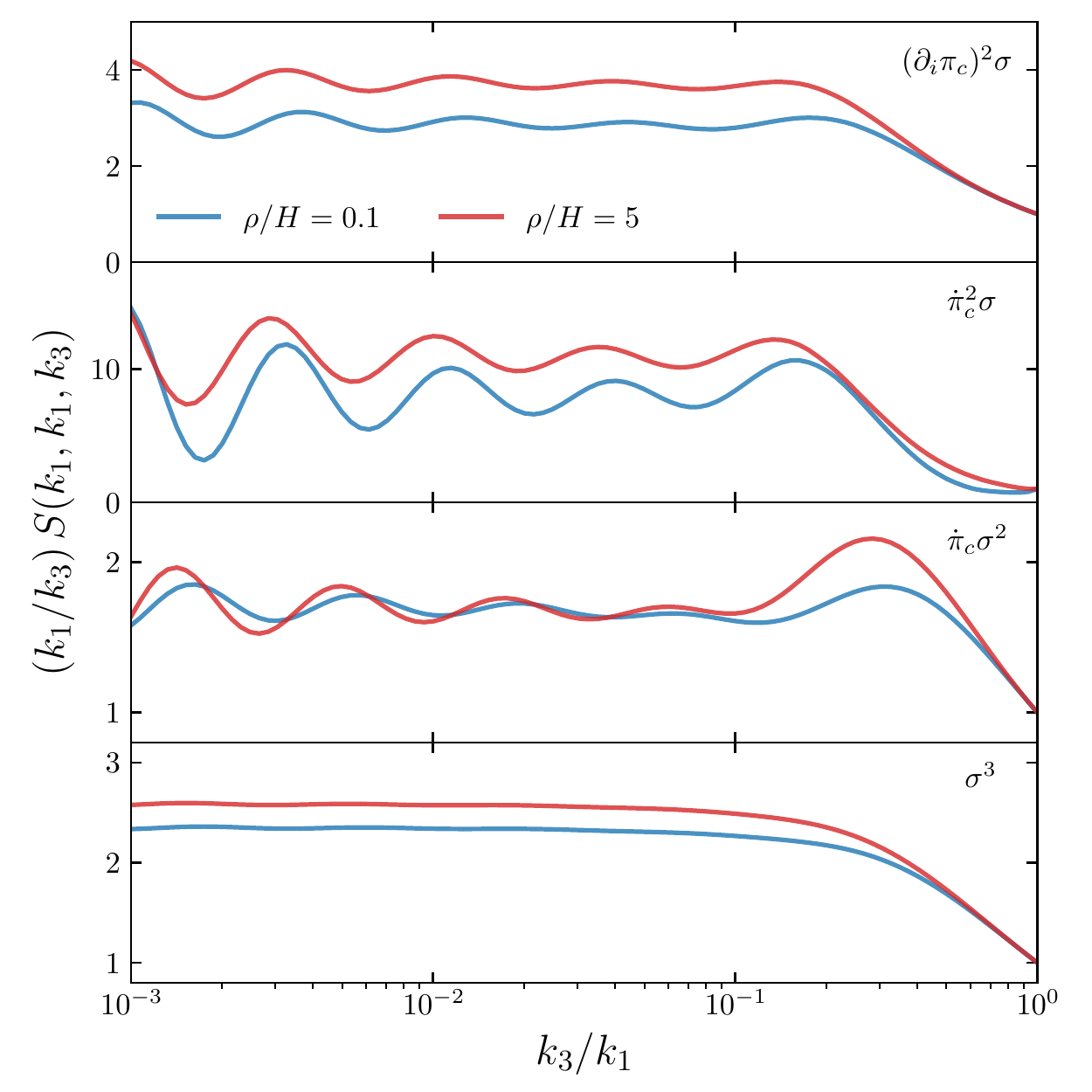}
    \caption{Squeezed limit of the shape functions for the four cubic interactions considered in Eq.~(\ref{eq: pi-sigma theory}) in the isosceles-triangle configuration $k_1=k_2$, for $c_s=0.1$ and $\mu_{\text{eff}} = 5$ varying the quadratic mixing $\rho/H= \textcolor{pyblue}{0.1}, \textcolor{pyred}{5}$, corresponding to \textcolor{pyblue}{weak} and \textcolor{pyred}{strong} mixings, respectively. For illustration purposes and to highlight the cosmological collider signal frequency, we have normalized the amplitudes of the signals to unity in the equilateral configuration $k_1=k_2=k_3$.
    \label{fig: CC signal weak/strong mixing}}
\end{figure}

\vskip 4pt
\paragraph*{\bf Cosmological Colliders at Strong Mixing.} At tree level, each cubic interaction in Eq.~(\ref{eq: pi-sigma theory}) gives an independent contribution to the bispectrum, and therefore can be treated separately. Following standard conventions~\cite{Babich:2004gb}, we define the shape function $S$ such that\footnote{A prime on a correlator indicates that we have dropped the momentum conserving delta function $(2\pi)^3 \delta^{(3)}(\bm{k}_1 + \bm{k}_2 + \ldots)$.}
\begin{equation}
    \braket{\zeta_{\bm{k}_1} \zeta_{\bm{k}_2} \zeta_{\bm{k}_3} }' \equiv \frac{(2\pi)^4}{(k_1k_2k_3)^2} \, \Delta_\zeta^4 \, S(k_1, k_2, k_3) \,,
\end{equation}
where $\Delta_\zeta^2 = \frac{k^3}{2\pi^2}\braket{\zeta_{\bm{k}} \zeta_{-\bm{k}}}'$ is the dimensionless power spectrum of $\zeta$. 
The squeezed limit $k_3 \ll k_1\simeq k_2$ is often described as a clean probe of the inflationary field content~\cite{Chen:2009zp, Noumi:2012vr, Arkani-Hamed:2015bza}, notably with oscillations of the type $S \sim (k_3/k_1)^{1/2} \cos [\mu \log(k_3/k_1)+\varphi]$ revealing the spontaneous production, by the expansion of the Universe, of a pair of scalar particles of masses $m/H=\sqrt{\mu^2+9/4}$. In Fig.~(\ref{fig: CC signal weak/strong mixing}), we compare these cosmological collider signals, obtained by numerically solving Eqs.~(\ref{eq: 2pt time evolution}, \ref{eq: 3pt time evolution}), for different constant quadratic mixing strengths, and for each cubic interaction in \eqref{eq: pi-sigma theory}. The most distinctive feature is that the frequency of the oscillations is set by $\mu_{\text{eff}}^2 = m_{\text{eff}}^2/H^2 - 9/4$, with $m^2_{\text{eff}} = m^2 + \rho^2$ playing the role of the effective mass for $\sigma$ both at early and late times \cite{Castillo:2013sfa, An:2017hlx, Iyer:2017qzw, CosmoFlow}. The appearance of the quadratic mixing in $m_{\text{eff}}$ can be interpreted as the result of resumming an infinite number of quadratic mixing insertions in the propagators of both fields, when using an interaction scheme where quadratic mixings are treated perturbatively. At strong mixing, the propagation of $\sigma$ is affected by the surrounding $\pi_c$ medium that interacts with it, leading to a self-energy correction. This is analogous to the electron self-energy correction in quantum electrodynamics due to its interaction with the photon \cite{Weinberg:1995mt, Peskin:1995ev}, the difference being that in our case such resummation occurs at tree level because Lorentz invariance is spontaneously broken. 

Focusing on a single property of the signal like its frequency, one could wrongly interpret a future detection as the discovery of a new weakly mixed massive particle, whereas this may well correspond to the signature of a light particle, albeit strongly mixed to the curvature perturbation. Our approach precisely allows one to break such degeneracies by providing complete predictions---covering the frequency, amplitude and phase of the cosmological colliders, as well as contaminations from equilateral shapes---for any theory. This enables one to explore the full range of possible signals without working under the lamppost of analytical tractability. Moreover, the cosmological flow approach also enables one to reveal the dynamics of fluctuations, as we show in \href{https://github.com/deniswerth/Cosmological-Collider-Flow}{\textcolor{purple}{\textsf{movies}}}\footnote{\href{https://github.com/deniswerth/Cosmological-Collider-Flow}{https://github.com/deniswerth/Cosmological-Collider-Flow}} displaying how the cosmological collider signals are built differently as inflation proceeds, in theories with weak and strong mixing.
This exemplifies how our method provides a powerful guide for physicists to test their theoretical understanding.


\vskip 4pt
\paragraph*{\bf Cosmological Colliders with Features.} Focusing on inflationary fluctuations, the background dynamics is encoded in the time dependence of $H_{\sf{ab}}, H_{\sf{abc}}, \ldots$. Here, we highlight the consequences of a time-dependent mixing on the cosmological collider signal. For definiteness, we consider $\rho(t) = \rho_0 (a/a_0)^{-n} \sin[\omega_c (t-t_0)]$ with $n \geq 0$.
The situation $n=0$ is relevant where the continuous shift symmetry of the Goldstone boson is broken to a discrete subgroup, see, e.g., \cite{Chen:2008wn, Flauger:2009ab, Flauger:2010ja, Chen:2010bka, Behbahani:2011it}, while $n=3/2$ is representative of, e.g., background trajectories that undergo a sudden turn in field space, after which a massive field relaxes to its minimum subject to underdamped oscillations with frequency $ \omega_c$.
We neglect any time variation of the Hubble parameter $H$, of the mass of $\sigma$, and of the cubic interaction strengths in the second line of \eqref{eq: pi-sigma theory}, as our purpose is to concentrate on what is imposed by symmetries. For simplicity, we also set $c_s=1$ and focus first on weak mixing $\rho_0<H$. The time-dependent mixing induces scale-dependent features in all correlation functions. For a fixed overall scale, the shape dependence of the cosmological collider depends on the cubic interaction that is considered. For those with constant strengths, the signal is conventional with frequency set by $\mu$~\cite{Chen:2022vzh}. Instead, the cubic interactions dictated by the quadratic mixing are intrinsically time dependent and have a special status, with $\dot{\rho}\,\pi_c\dot{\pi}_c\sigma$ dominant for rapid oscillations with $\mu_c \equiv \omega_c/H > 1$.

\begin{figure}
    \centering    \includegraphics[width=1\columnwidth,trim={0pt 6pt 0pt 0}]{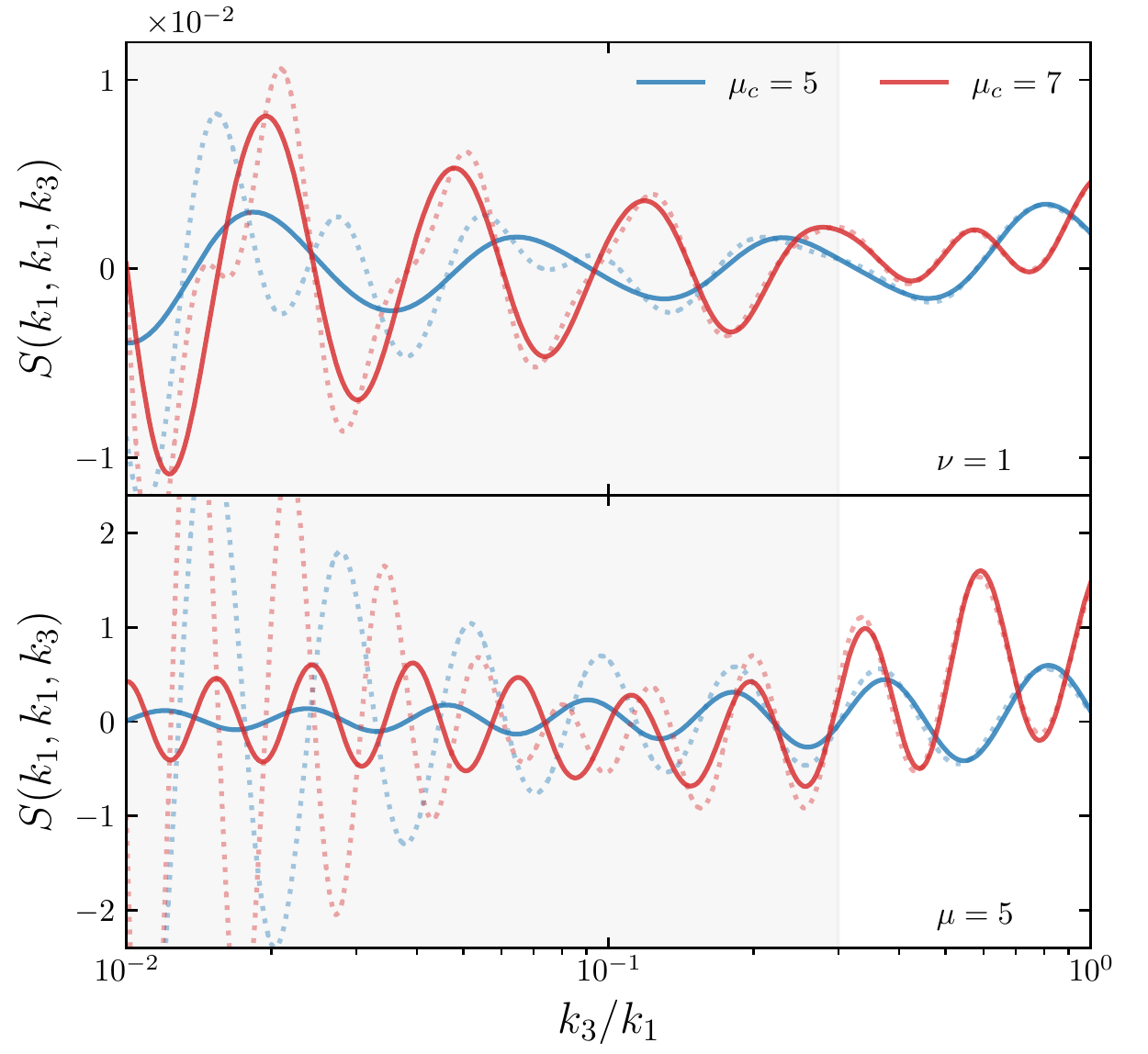}
    \caption{Shape dependence of the observable bispectrum (\textit{solid line}) dictated by the time-dependent mixing $\rho(t)$, in the isosceles-triangle configuration $k_1=k_2$ at a fixed scale $\log(k_3/k_0)=2$ where $k_0=(a H)(t_0)$, for a 
    light field $\nu=1$ (\textit{top}) and a heavy field $\mu=5$ (\textit{bottom}), varying the background frequency $\mu_c = \textcolor{pyblue}{5}, \textcolor{pyred}{7}$, and setting $n=0$. 
    The observable signal is effectively reconstructed from the computed one (\textit{dotted line}) upon meticulously subtracting the consistency relation 
    $P_\zeta(k_\textrm{S})P_\zeta(k_3) \dd \log[k_\textrm{S}^3 P_\zeta(k_\textrm{S})]/ \dd \log k_\textrm{S}$ with $\bm{k_\textrm{S}}=\bm{k}_1+\bm{k}_3/2$,
    inferred from the power spectrum $\braket{\zeta_{\bm{k}} \zeta_{-\bm{k}}}' = P_\zeta(k)$. The shaded region corresponds to the squeezed limit where the templates (\ref{eq: CC signals with features}) are valid. We have chosen $\rho_0/H=0.1$ to respect the current bounds on the power spectrum~\cite{Planck:2018jri}.}
    \label{fig: CC signal light heavy} 
\end{figure}

We show in Fig.~(\ref{fig: CC signal light heavy}) the corresponding shape dependence of the bispectrum for a light and heavy field $\sigma$, for $n=0$ and different background frequencies $\mu_c$.
These shapes are directly measurable in cosmological data as they already take into account the rescaling of the background by the long $\zeta$ mode \cite{Tanaka:2011aj, Pajer:2013ana}, yet leading to striking behaviors. 
In the squeezed limit, we find 
\begin{widetext}
\begin{equation}
    S(k_1, k_1, k_3) =
    \begin{cases}
      \left(\frac{k_3}{k_1}\right)^{1/2+n - \nu} \mathcal{A}\cos\left( \mu_c \log\left(\frac{k_3}{k_1}\right) + \varphi\right)\,, \quad \quad \text{for} \quad (\nu ,n) \neq (3/2,0)\,, \quad & \text{(light)}  \\
      \left(\frac{k_3}{k_1}\right)^{1/2+n}\left[\mathcal{A}_+ \cos\left((\mu+\mu_c)\log\left(\frac{k_3}{k_1}\right) + \varphi_+\right) + \mathcal{A}_- \cos\left((\mu-\mu_c)\log\left(\frac{k_3}{k_1}\right) + \varphi_-\right)\right]\,,  & \text{(heavy)} 
    \end{cases} 
    \label{eq: CC signals with features}
\end{equation}
\end{widetext}
where $\nu = i\mu$. Because of the dilution of the feature, the power-law scaling acquires an additional $(k_3/k_1)^{n}$  suppression compared to the usual one~\cite{Chen:2009zp}. 
Depending on whether the field $\sigma$ is light ($m/H \leq 3/2$) or heavy ($m/H \geq 3/2$), the cosmological collider signal is either dictated by the background frequency $\mu_c$ or presents modulated frequencies in $\mu \pm \mu_c$, respectively. The observed beatings follow from mode mixing between two features: classical oscillations of the coupling and quantum oscillations of the massive field on super-Hubble scales.
Remarkably, we find numerically that the templates \eqref{eq: CC signals with features} hold in the strong mixing regime. In particular, for light fields with $\nu > 1/2 + n$, we find that the oscillations modulate a \textit{growing} envelope in the squeezed limit, which is expected to generate distinctive oscillations in the scale-dependent galaxy bias.

\vskip 4pt
\paragraph*{\bf Conclusions.} The physics of inflation is phenomenologically rich and complex. In this Letter, we have presented a systematic framework for computing primordial correlators in any theory, focusing for definiteness on scalar degrees of freedom. Although massive fluctuations decay during inflation, their presence leaves a smoking-gun imprint in the squeezed limit of the observable bispectrum. Known as the cosmological collider signal, it offers a thrilling opportunity to identify new particles, in the same way as resonances for ground-based colliders. We demonstrated the power of our approach by computing cosmological collider signals in theories with strong and time-dependent mixings, and for various kinds of cubic interactions. These theoretically motivated scenarios present challenges that analytical methods are unable to address.

The cosmological flow provides the means for exploring and understanding inflationary physics in full generality, bridging the gap between theories and observations.
To ensure its accessibility, our new computational approach is available as an open-source numerical tool~\cite{Werth:2024aui}.
As natural extensions, we plan to incorporate spinning fields and higher-order correlation functions, as well as consider loops.



\vskip 4pt
\begin{acknowledgments}
We are grateful to 
Xingang Chen,
Paolo Creminelli,
Sadra Jazayeri,
David Mulryne,
Toshifumi Noumi,
Enrico Pajer,
David Seery and Xi Tong
for useful comments on a draft version of this Letter. We would like to especially thank David Mulryne and David Seery for enlightening conversations when this work was initiated. DW and SRP are supported by the European Research Council under the European Union's Horizon 2020 research and innovation programme (grant agreement No 758792, Starting Grant project GEODESI). LP acknowledges support from the Atracción de Talento grant 2019-T1/TIC15784 when this work was started, his work is now supported by the Spanish Research Agency (Agencia Estatal de Investigación) through the Grant IFT Centro de Excelencia Severo Ochoa No CEX2020-001007-S, funded by MCIN/AEI/10.13039/501100011033. This article is distributed under the Creative Commons Attribution International Licence (\href{https://creativecommons.org/licenses/by/4.0/}{CC-BY 4.0}).
\end{acknowledgments}

\bibliography{Bibliography}

\end{document}